\documentclass[reprint,aps,prl,showpacs,superscriptaddress]{revtex4-1}
\usepackage{graphicx,epsfig}
\usepackage{epstopdf}
\usepackage{bm}

\begin{document}

\title{Dissipation of Quasiclassical Turbulence in Superfluid $^4$He}
\author{D. E. Zmeev}
\affiliation{School of Physics and Astronomy, The University of Manchester, Manchester M13 9PL, UK}
\affiliation{Department of Physics, Lancaster University, Lancaster LA1 4YB, UK}
\author{P. M. Walmsley}
\affiliation{School of Physics and Astronomy, The University of Manchester, Manchester M13 9PL, UK}
\author{A. I. Golov}
\affiliation{School of Physics and Astronomy, The University of Manchester, Manchester M13 9PL, UK}
\author{P. V. E. McClintock}
\affiliation{Department of Physics, Lancaster University, Lancaster LA1 4YB, UK}
\author{S. N. Fisher}
\altaffiliation{Deceased 4 January 2015.}
\affiliation{Department of Physics, Lancaster University, Lancaster LA1 4YB, UK}
\author{W. F. Vinen}
\affiliation{School of Physics and Astronomy, University of Birmingham, Birmingham B15 2TT, UK}
\date{\today}

\begin{abstract}
We compare the decay of turbulence in superfluid $^4$He produced by a moving grid to the decay of turbulence created by either  impulsive spin-down to rest or by intense ion injection. In all cases the vortex line density ${\cal L}$ decays at late time $t$ as ${\cal L} \propto t^{-3/2}$. At temperatures above 0.8\,K, all methods result in the same rate of decay. Below 0.8\,K, the spin-down turbulence maintains initial rotation and decays slower than grid turbulence and ion-jet turbulence. This may be due to a decoupling of the large-scale superfluid flow from the normal component at low temperatures, which changes its effective boundary condition from no-slip to slip. 

\end{abstract}

\maketitle

Turbulence is a common state of flow in classical fluids, with great importance from atmospheric systems to aircraft design. So far, satisfactory understanding is only achieved for homogeneous and isotropic turbulence (HIT) \cite{K41,HIT}. HIT can be approximately obtained in the wake of a flow past a grid \cite{GridTurbulence, Isaza2014}, although it might still be strongly modified by the container geometry \cite{CommentMultiplicity, Huisman2014}.
Grid turbulence in superfluid $^4$He was obtained  \cite{Smith93,Stalp99,Skrbek2000}, but not at temperatures below 1\,K due to technical difficulties. Yet,  
the low-temperature regime enjoys a special interest, as the thermal excitations (the normal component) are essentially absent. Turbulence of the superfluid is made of a chaotic motion of tangled topological defects of the superfluid order parameter field -- quantized vortices --
each carrying the same circulation equal to the ratio of the Planck's constant to the mass of a $^4$He atom:  $\kappa=h{\rm \,}m^{-1}$. It is called Quantum Turbulence (QT), as it is essentially a macroscopic quantum phenomenon. QT decays even at the lowest temperatures, but the mechanisms of dissipation in superfluid $^4$He -- thought to be the radiation of phonons by Kelvin waves (perturbations of vortex lines) with wavelength $\sim 10^{-7}$\,cm~\cite{Vinen01} and also of small ballistic vortex loops that can carry energy away \cite{Svistunov1995,BarenghiRings,NemirovskyRings,KursaRings} -- only operate at very small length scales.  Existing theories \cite{LNR2007,KSPRB2008,KS2008,KSJLTP,SoninPRB2012,Barenghi14,Boue2015} of QT decay are applicable to homogeneous isotropic QT (HIQT), for which only sparse experimental data are available in the interesting ultra-low temperature limit.

In this Letter we report the best-yet realization of HIQT in the $T \rightarrow 0$ limit. We measure the
 free decay of grid turbulence and compare the results with both theory and experiments using other methods, thereby gaining valuable insights into the underlying processes.

 When QT is generated by large-scale flow, on length scales much greater than the mean intervortex distance $\ell_q={\cal L}^{-1/2}$, where ${\cal L}$ is the length of vortex lines per unit volume, then the energy is predominantly contained in flow at the largest length scales $\gg \ell_q$. In this case QT is called {\it quasiclassical}~\cite{Vinen2002,Walmsley14}, as quantization of vorticity becomes unimportant, and the coarse-grained velocity field is expected to obey the Euler equation.  It is believed that this energy cascades towards the smaller length scales via a classical hydrodynamic cascade, followed, at length scales $\leq \ell_q$, by a `quantum cascade' that involves reconnections and Kelvin waves on discrete vortex lines. Existing theories \cite{LNR2007, KSPRB2008, KSJLTP, SoninPRB2012, Barenghi14} of these processes in HIQT all assume that the dominant contribution to $L$ is at quantum mesoscales $\sim \ell_q$, but they differ in detail.
 For self-similar flows, assuming that
 the rate of dissipation of flow energy per unit mass, $\cal{E}$, only depends on ${\cal L}$ and $\kappa$, dimensional considerations demand
\begin{equation}
\dot{\cal{E}} = - \zeta\kappa^3 {\cal L}^2.
\label{nu-prime}
\end{equation}
Here, the `non-dimensional effective kinematic viscosity' $\zeta \sim 1$ (the more conventional `effective kinematic viscosity' is $\nu' \equiv \zeta\kappa$) \cite{Vinen2000,Vinen2002}. At medium temperatures $1.0 \lesssim T \lesssim 1.6$\,K, it reflects the dissipation through the interaction of vortices with thermal excitations (expressed through the `mutual friction parameter' $\alpha(T)$), while in the limit $T \rightarrow 0$ ($ T \lesssim 0.5$\,K) it  characterizes the efficiency of the tangle of vortex lines in maintaining the energy cascade down to the dissipative length scale.
 As there is no microscopic derivation of (\ref{nu-prime}), it remains unclear whether the value of $\zeta$ is the same for HIQT of any spectrum, or depends on the type of flow. For instance, $\zeta = 0.08$ was measured \cite{Walmsley08} for Vinen (`ultraquantum', i.\,e. without flow at classical length scales $> \ell_q$) QT at $T \rightarrow 0$, while the analysis of the decay of QT generated by spin-down at $T \rightarrow 0$ apparently revealed $\zeta \approx 0.003$ \cite{Walmsley07}. The latter was heralded as  evidence for the poor efficiency of the energy cascade in quasiclassical QT due to the `bottleneck'  between the classical and quantum lengthscales \cite{LNR2007}. However, recent experiments in a rotating container revealed vanishing traction by the container walls on turbulent superfluid $^3$He at low temperatures when $\alpha < 10^{-3}$, resulting in a  long-lived rotating state \cite{Hosio2013}. This cast doubt on the interpretation of $^4$He spin-down turbulence as being HIQT \cite{Walmsley07} and pushed for new experiments with truly HIQT.

 Thus, the goal of this work was to measure and compare the decay rates of different types of turbulent flow, including those generated by a towed grid and impulsive spin-down, in a broad range of temperatures. To determine the value of $\zeta$, one has to know  both ${\cal L}$ and $\dot{\cal{E}}$ in (\ref{nu-prime}). With our technique of free decay, the injected energy flux, $-\dot{\cal{E}}$, is controlled  by the size of the largest energy-containing eddy and its lifetime.
In fact, (\ref{nu-prime}) with a meaningful $\zeta$ can only be applied for homogeneous turbulence while, for bound inhomogeneous flows, only an integral rate of energy dissipation can be measured together with some averaged  value of vortex line density. We will hence assume that (\ref{nu-prime}) relates average $\dot{\cal{E}}$ and ${\cal L}$ through some  integral $\zeta$.

The energy per unit mass of helium in the energy-containing eddies with velocity amplitude $u$ is ${\cal{E}} = \xi u^2$, where $\xi \lesssim 1/2$. Their length scale $\lambda$ is limited by the  container size $d$, $\lambda = \beta d$, where $\beta \sim 1$. We assume that, as in classical turbulence, this energy is released within the lifetime $\tau$ of order the turn-over time $\sim \lambda u^{-1}$, i.\,e. $\tau = \theta \lambda u^{-1}$, where $\theta \sim 1$.
In the quasi-steady regime, the energy flux fed into the cascade is hence $-\dot{\cal{E}} = {\cal{E}}\tau^{-1}$ or
\begin{equation}
-2\xi u \dot{u} = \xi u^3 \theta^{-1} \beta^{-1} d^{-1}.
\label{E-dot}
\end{equation}
Its solution at late time $t$ is
\begin{equation}
{\cal{E}}(t) = 4\xi \theta^2\beta^2d^2t^{-2},
\label{E-t}
\end{equation}
\begin{equation}
\tau(t) = t/2.
\label{tau-t}
\end{equation}
After plugging (\ref{E-t}) into (\ref{nu-prime}), we arrive at
\begin{equation}
{\cal L}(t)\sim A d (\kappa t)^{-3/2},
\label{3/2}
\end{equation}
where $A \equiv (8\xi)^{1/2}\theta \beta \zeta^{-1/2} \sim 1$.
This is the ${\cal L}\propto t^{-3/2}$ free decay that was observed in many experiments \cite{Stalp99,Bradley2006,Walmsley07,Walmsley08} and numerical simulations \cite{Risto}.

\begin{figure}
\includegraphics[width=6cm]{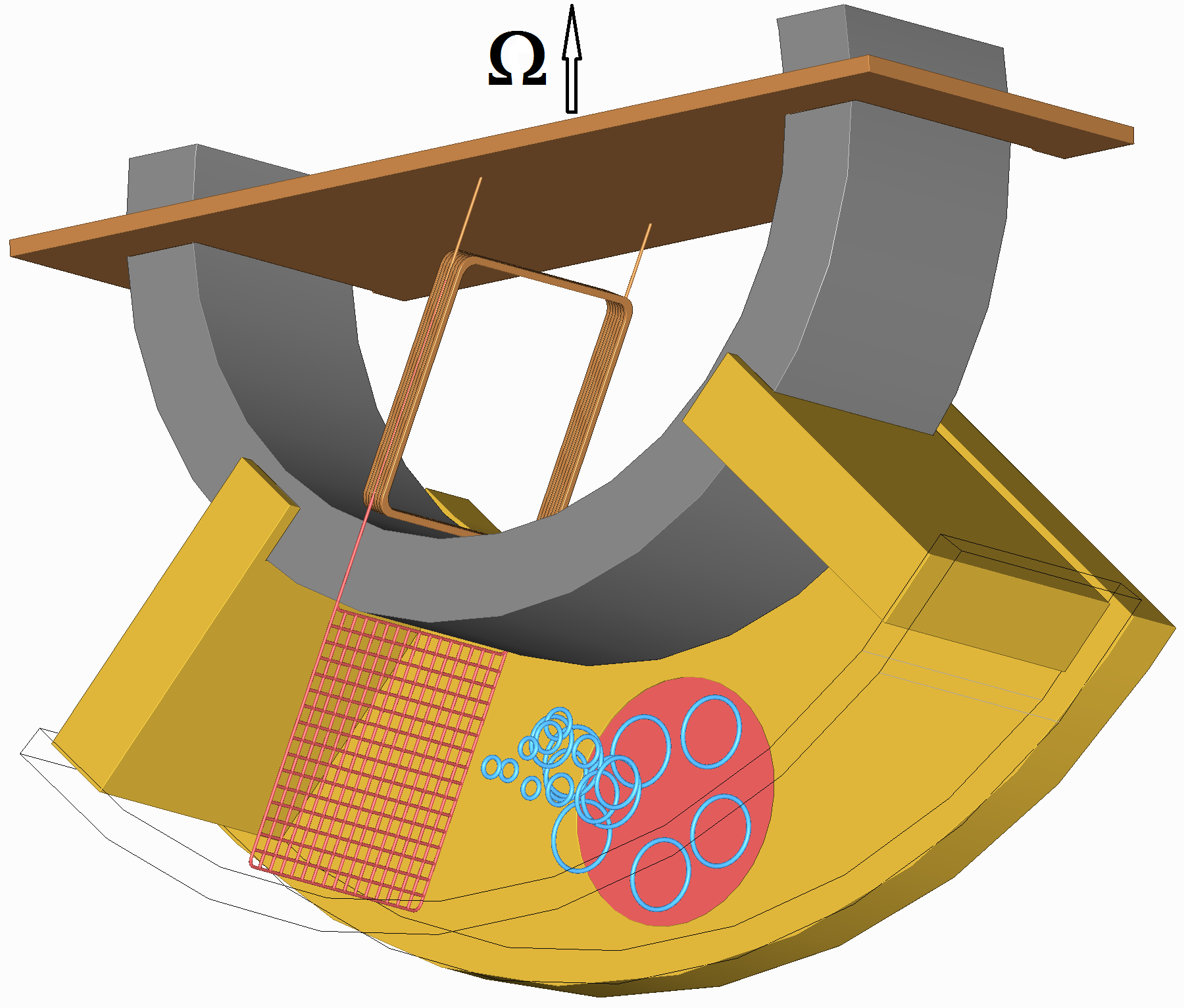}
\caption{Experimental setup \cite{Zmeev14}. The front and bottom walls of the channel are not shown. Blue circles depict charged vortex rings (CVRs,  not to scale), used to probe QT. CVRs propagate from the injector (not shown) at the front wall to collector inside a hole (shown by red circle) in the back wall. The assembly could be rotated about the vertical axis ($\bf \Omega$).}
\label{channel}
\end{figure}

Our experiments were conducted in ultra-pure \cite{purity,purity2} $^4$He at pressure $0.1$\,bar filling the volume shown in Fig.~\ref{channel}: a 90\,$^\circ$ section of an earthed annular channel with an inner wall radius of curvature equal to 2.75\,cm and of rectangular cross-section with sides $d_h=1.8$\,cm\,(horizontal) and $d_v=1.7$\,cm\,(vertical). A brass grid (1.5\,cm$\times$1.5\,cm) could be electromagnetically driven at a constant velocity from one end of the channel to the other. The operating principle of the device is described elsewhere \cite{Zmeev14} while the technique of measuring the density and polarization of vortex tangles using negative ions is detailed in section 1 of Supplemental Material (SM) \cite{SM}.

\begin{figure}
\includegraphics[width=7cm]{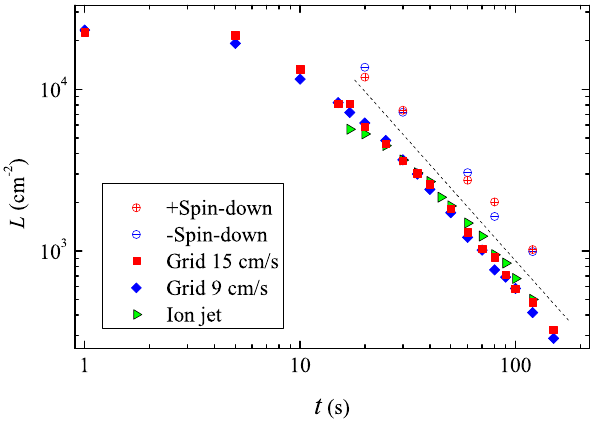}
\caption{ Decay of vortex line density ${\cal L}(t)$ for turbulences, generated by different means: `+Spin-down' -- spin-down from $\Omega=$+1.5\,rad\,s$^{-1}$,  `-Spin-down' -- spin-down from $\Omega$=-1.5\,rad\,s$^{-1}$, `Grid 15\,cm\,s$^{-1}$' -- grid with $m_g = 3$\,mm and $v_g=15$\,cm\,s$^{-1}$, `Grid 9\,cm\,s$^{-1}$' -- grid with $m_g = 3$\,mm and $v_g=9$\,cm\,s$^{-1}$, `Ion jet' -- injection of negative ions at a current of 700\,pA lasting for 100\,s. Dashed line shows the $t^{-3/2}$ dependence.  
$T=80$\,mK.}
\label{L_vs_t}
\end{figure}

We investigated the decay of turbulence generated by three different methods: towing a grid at velocity $v_g \sim 10$\,cm\,s$^{-1}$ through the channel \cite{OscillatingGrid}, impulsive spin-down from uniform rotation at angular velocity $\Omega \sim 1$\,rad\,s$^{-1}$ to rest, and injection of  electric current for long periods of time ($\sim 1$\,nA through voltage $\sim 100$\,V for $\sim100$\,s). Each resulted in well-developed quasi-classical turbulence in a wide range of length scales (the length scales and corresponding effective Reynolds numbers are tabulated in section 2 of \cite{SM}). After generation, ${\cal L}(t)$ was probed with a pulse of ions after a delay $t$. Each realisation was probed only once to avoid distortion of the turbulent flow by the probing pulses.
For each method, we forced QT sufficiently hard, that the late-time decay was the same, independent of the intensity of forcing
 (e.\,g. if $v_g \gtrsim 5$\,cm\,s$^{-1}$).
 In the experiments with grid turbulence the values ${\cal L}(t)$  at late times did not depend on how many times in succession (1, 2, 3 or 10) the grid was towed through the channel, nor did it depend on the grid mesh sizes $m_g$ used (0.75\,mm and 3\,mm). In the experiments with rotation the grid was parked at one end of the channel.

\begin{figure}
\includegraphics[width=7cm]{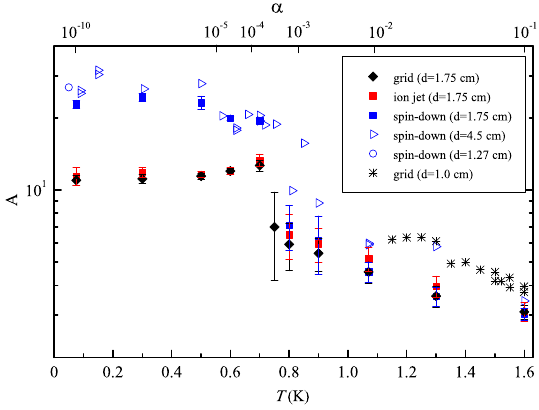}
\caption{ The values of the fitting parameter $A={\cal L}(t)d^{-1}(\kappa t)^{3/2}$ vs. temperature (values of the mutual friction parameter $\alpha (T)$ are shown at the top). Measurements by Stalp {\it et al.} \cite{Stalp2002} are shown by asterisks for comparison. }
\label{AvsT}
\end{figure}

For all temperatures and all methods of turbulence generation, after a method-specific transient process of duration $\lesssim 10$\,s, the decays of vortex line density followed ${\cal L} \propto t^{-3/2}$,  as shown in Fig.~\ref{L_vs_t}.
We fitted them to (\ref{3/2}) \cite{OriginTime} for time $t$ between 30\,s and 200\,s, and the resulting values of $A(T)$ (using $d=(d_h+d_v)/2 = 1.75$\,cm) are plotted in Fig.\,\ref{AvsT}.
We also compare these with the experimental values of $A(T)$ for grid turbulence (square channel, $d=1.27$\,cm) \cite{Stalp99,Stalp2002} and spin-down turbulence (cubic cell, $d=4.5$\,cm \cite{Walmsley07} and rectangular cell, $d=1.27$\,cm \cite{Walmsley14}).
One can see that at temperatures above 0.8\,K, corresponding to $\alpha >  10^{-3}$, the values of $A(T)$ for all methods of turbulence generation in our container agree with each other and also, within their scatter, with previous experiments. However, at $T< 0.8$\,K  $A(T)$  approaches  either of two zero-temperature limits: $A(0)\approx 11$ for both the ion-jet and grid-generated turbulence, while $A(0)\approx 23$ for the spin-down turbulence.
We would thus conclude that at $T>0.8$\,K the late-time turbulence is the same  whatever the initial flow, i.\,e. approximatelly isotropic and homogeneous. This implies that
the leftovers of the initial flow pattern (say, rotation following spin-down) disappear within less than 30\,s. But at lower temperatures,   the spin-down turbulence is different from that for other methods at all times; this might be explained by our observation that the memory of initial rotation is retained during the late-time decay \cite{SM} -- presumably in the form of a vortex tangle, rotating at angular velocity $\sim 0.1$\,rad\,s$^{-1}$ near the vertical axis of the cell, that preserves some of the initial angular momentum.

During the transient following the spin-down of a  rectangular cell, much of the fluid's initial angular momentum is transfered to the walls through pressure fluctuations from large eddies, eventually creating turbulence with a broad distribution of length scales. At late times, when, as we suppose, the remains of that angular momentum survive only  near the axis, these  pressure fluctuations at walls (`pressure drag') become inefficient, and only the traction at the walls (`frictional drag') exerts torque. If at $T<0.8$\,K this traction becomes too small to reduce the remaining angular momentum within the decay time, the effective boundary conditions (BC) become of the `slip' type. Let us discuss two different origins of traction: the viscosity of the normal component and vortex pinning.

For laminar flow, the relaxation time for coupling between the superfluid and a stationary normal component is $\sim [\alpha(T) \kappa {\cal L}]^{-1}$. With decreasing temperature, it rapidly increases  and should be compared to the lifetime of energy-containing eddies (\ref{tau-t}): the cross-over from the limit of coupled to uncoupled components would thus be expected  at $\alpha \sim 2[\kappa t{\cal L}(t)]^{-1}$. For typical ${\cal L}(t) \sim 10^4$\,cm$^{-2}$ at $t \sim 20$\,s (as in Fig.\,\ref{L_vs_t}), this corresponds to $\alpha(T) \sim 10^{-2}$, i.\,e. to $T\sim 1.1$\,K. However, in a turbulent state, the locally-enhanced density of vortex lines near walls might enhance the mutual friction force, hence allowing the cross-over to occur at smaller values of $\alpha(T)$. Furthermore, as the mechanical forcing is expected to affect the large-scale superfluid and normal flow in a similar manner, these flows could be generated nearly fully-coupled from the outset; this may further ease the condition for coupling and allow the cross-over to `slip' BC to occur at a lower temperature. Note that rotation of superfluid $^3$He was also found to decouple from container walls when $\alpha \lesssim 10^{-3}$  \cite{Hosio2011,Hosio2013,EltsovPNAS2014}.

With numerous vortex lines terminated at the container walls, a tangential flow experiences an effective friction due to the pinning of these lines \cite{Adams}. This force depends on the roughness of the surface, density of vortex lines as well as the lines' dynamics -- such as the frequency of reconnections (that facilitates effective depinning of lines) and tension in the presence of developed Kelvin waves. We can give a conservative estimate of the upper limit on this force per unit area, $F_p < f {\cal L}$, by assuming that all lines are strongly pinned \cite{Roughness} and pull in the direction of tangential flow with force equal to their line energy, $f = \frac{\rho\kappa^2}{4\pi}\ln\left(\frac{\ell_q}{a_0}\right) \approx 1.5$\,pN, where $a_0 \sim 1{\rm \,\AA}$ is the radius of the vortex core and $\rho$ is the density of helium. Such a force would remove the angular momentum in a cell of square cross-section with side $d_h=1.8$\,cm, initially rotating at $\Omega_0=1.5$\,rad\,s$^{-1}$, in $\sim \frac{\rho d^2\Omega_0}{24 f {\cal L}}$, which is $\sim 20$\,s for ${\cal L}(t)\sim 10^4$\,cm$^{-2}$ at $t=20$\,s.
While this relaxation time is indeed comparable with the decay time, the force in a realistic weakly-polarized tangle should be much weaker. Furthermore, reconnections of pinned vortex lines can play an important role in promoting their creep from one pinning site to another \cite{ZievePRB2012}; this effective reduction of the friction force is believed to be facilitated by the enhanced amplitude of Kelvin waves on the scale of wall roughness -- which are expected to rapidly grow in size when $\alpha < 10^{-3}$, i.\,e. damping due to mutual friction becomes negligible \cite{KSPRB2008}. Lastly, because of frequent reconnections, the torque cannot extend much beyond one mean inter-vortex distance $\ell_q$. Hence,  only a vanishingly small shear stress can be sustained by the tangle, and it will be impossible to exert sufficient torque on the rotating core far from the container walls.


It is thus not surprising that at $T<0.8$\,K, the  decoupling of the superfluid component from the container at large length scales and time scales of order the decay time results in long-lived rotation far from walls.
As in the classical case \cite{RotatingTurbulence}, this residual rotation should slow down the cascade of energy to smaller eddies  and thus  increase the value of $\theta$.
Hence, according to (\ref{3/2}), this can explain the fact that, at the same decay time $t$, the vortex line density ${\cal L}(t)$ is higher for spin-down turbulence than for grid turbulence.  It is also comforting to see that spin-down turbulence in containers with three different $d$ all returned similar zero-temperature $A\equiv {\cal L}(\kappa t)^{3/2}d^{-1}$ in Fig.\,\ref{AvsT} (different blue symbols), as predicted by (\ref{3/2}).

Let us now discuss the possible effect of BC on the decay rate of grid turbulence. Far from walls, the dynamics of the superfluid eddies (whether coupled to the low-viscosity normal component at $T \gtrsim 1$\,K or decoupled from the vanishing normal component at $T \lesssim 1$\,K) at classical length scales is expected to be identical \cite{Vinen2000}.
 However, this is not the case for the energy-containing eddies in a container, because they are affected by walls.
 No-slip BC would speed-up the breakdown of eddies through the diffusion of vorticity via eddy viscosity and thus decrease the parameter $\theta$ relative to its bulk value for eddies of the same size; while slip BC might actually increase the value of $\theta$. The effective size of the largest eddies in a container might also be greater for slip BC, which will be reflected in a larger value of $\beta$. Either effect could thus explain an increase of the parameter $A \propto \beta \theta \zeta^{-1/2}$ if BC becomes of slip type below 0.8\,K -- even if the effective kinematic viscosity $\zeta(T)$ stayed the same.

As the values  of the parameters $\xi$, $\beta$ and $\theta$ for a container of particular shape and BC are unknown, it is impossible to determine the accurate value of $\zeta$ from  $A$.
Stalp {\it et al.} \cite{Stalp99} introduced an approach, in which they assumed that  the  energy spectrum in the space of wavenumbers $k$  is meaningful and equal to the Kolmogorov spectrum $E_k=C\epsilon^{2/5}k^{-5/3}$ (with $C\approx 1.5$) all the way down to the the cut-off wavenumber $k_1 \sim d^{-1}$.
  In section 3 of SM \cite{SM}  we show that these assumptions are unrealistic, and one hence cannot expect accurate  values of $\zeta$ from this approach.
Yet, we quote its result for $T=0$: for slip BC (for which $k_1 = \pi/d$), the value $A(0)=11$  for grid turbulence (Fig.\,4) would correspond to $\zeta(0) \approx 0.08$.
This agrees well with values $\zeta (0) =$ 0.08--0.09  measured experimentally \cite{Walmsley08,CommentUltraquantum}  and $\zeta(0) =$ 0.06--0.10 calculated numerically \cite{Tsubota2000,Kondaurova2014} for Vinen  QT, in which classical degrees of freedom are not excited. It seems the same bulk parameter $\zeta(T)$ characterizes the efficiency of quantum cascades in HIQT for different spectra, thus suggesting that there is no bottleneck between the classical and quantum cascades

To conclude, by towing a grid through superfluid helium in the zero-temperature limit, we have produced the best-yet realization of quasiclassical HIQT filling a container, and measured its decay rate.  The low-temperature decay of HIQT follows the law ${\cal L} \propto t^{-3/2}$, observed for all quasi-classical QT, but its decay is markedly faster than that of the turbulence generated by an impulsive spin-down to rest. The latter may be due to the change of the effective BC from no-slip to slip because of the loss of traction at the container walls below 0.8\,K. As a result, the spin-down flow  maintains rotation, which is responsible for the slowing-down of the decay of turbulence.

This work was supported through the Materials World Network program by the Engineering and Physical Sciences Research Council (Grant No. EP/H04762X/1). P. M. W. is indebted to EPSRC for the Career Acceleration Fellowship (Grant No. EP/I003738/1). We thank V. B. Eltsov and V. S. L'vov for fruitful discussions.


\begin{thebibliography}{99}
\bibitem{purity} $^4$He gas with $^3$He concentration  $2 \times 10^{-11}$ was obtained using the heat-flush technique \cite{purity2}.
\bibitem{OscillatingGrid} We have also investigated turbulence generated by the same grid oscillating  with amplitude $\sim 0.6$\,mm and frequency $\sim 15$\,Hz, 2\,cm away from the detector for typically 100\,s. ${\cal L}(t)$ was measured after the oscillations were stopped. Although turbulence produced in this way is initially highly inhomogeneous (with ${\cal L}$ decreasing with distance from the grid), at $t>100$\,s ${\cal L}(t)$ showed the same late-time decay as after a towed grid (Fig.\,2).
\bibitem{OriginTime} While the general form of the solution (\ref{3/2}) has form ${\cal L} \propto (t^*+t)^{-3/2}$, where $t^*$ is a constant of order several seconds, introducing another free parameter $t^*$ does not improve the accuracy of  determination of the parameter $A$ due to the presence of the transient processes of duration comparable with $t^*$. We thus only relied on fitting to (\ref{3/2}) at sufficiently late time $t \gg t^*$.
\bibitem{Roughness} We would estimate the roughness of the container walls to be on length scales $\lesssim 10$\,$\mu$, while the typical intervortex distance $\ell_q$ varied between $\sim 70$\,$\mu$ and $\sim 600$\,$\mu$.
\bibitem{CommentUltraquantum} We have also generated Vinen turbulence in this cell, using the method of tangling a small packet of charged vortex rings as in \cite{Walmsley08} and observed the decay ${\cal L} =1.6\times10^4{\rm \,s\,cm}^{-2} t^{-1}$, which is close to ${\cal L} =1.3\times10^4{\rm \,s\,cm}^{-2} t^{-1}$ observed in a cubic cell of 4.5\,cm side  \cite{Walmsley08}. The independence of the container size for this type of turbulence is indeed expected from theory: ${\cal L} \approx 1.2\kappa^{-1}\zeta^{-1}t^{-1}$ \cite{Walmsley08}.
\end{thebibliography}
\end{document}